# Thermal Stability and Phase Transformation of Conductive $\alpha$-(Al$_x$Ga$_{1-x}$)$_2$O$_3$/Ga$_2$O$_3$ Heterostructure on Sapphire Substrates

Botong Li[1,*], Shisong Luo[2,*], Jaeheon Jung[3,*], Bobby G. Duersch[4], Cheng Chang[2], Lucas Lau[2], Zonghao Zhang[2], Jianhua Li[5], Hunter Ellis[1], Imteaz Rahaman[1], Roy Byung Kyu Chung[3, a)], Kai Fu[1, a)], and Yuji Zhao[2, a)]

**AFFILIATIONS**

[1]Department of Electrical and Computer Engineering, University of Utah, Salt Lake City, UT 84112, USA

[2]Department of Electrical and Computer Engineering, Rice University, Houston, TX 77005, USA

[3]Advanced Materials Science and Engineering Department, Kyungpook National University, Daegu 41566, Republic of Korea

[4]Utah Nanofab Electron Microscopy and Surface Analysis Laboratory, The University of Utah, Salt Lake City, UT 84112, USA

[5]Shared Equipment Authority, Rice University, Houston, Texas 77005, USA

[*] Authors contribute equally to this work.

[a)] Authors to whom correspondence should be addressed: roy.b.chung@knu.ac.kr, kai.fu@utah.edu, and yuji.zhao@rice.edu

## Abstract

Thermal stability and phase transformation of conductive $\alpha$-(Al$_{0.16}$Ga$_{0.84}$)$_2$O$_3$/Ga$_2$O$_3$ heterostructure on sapphire substrates were investigated via in situ high temperature X-ray diffraction (HT-XRD), scanning electron microscopy (SEM), and atomic force microscopy (AFM). The conductive $\alpha$-(Al$_{0.16}$Ga$_{0.84}$)$_2$O$_3$/Ga$_2$O$_3$ heterostructure with fluorine (F) doping were growth by mist CVD on sapphire substrates, which achieved a Hall mobility of 28 cm$^2$/(V·s) and electron concentration of $1.4\times10^{20}$ cm$^{-3}$. The heterostructure exhibited thermal stability to ~550-575°C before transforming to $\beta$-(Al$_x$Ga$_{1-x}$)$_2$O$_3$/Ga$_2$O$_3$. The transformed $\beta$-Ga$_2$O$_3$ is mainly polycrystalline rather than a high-quality epitaxial phase. Reciprocal space mapping (RSM) results reveal that the edge dislocation density remains consistently higher than the screw dislocation density throughout the entire heating process, indicating that the crystalline imperfection in $\alpha$-Ga$_2$O$_3$ is dominated by in-plane mosaicist. After the phase transformation of $\alpha$ phase to $\beta$ phase, catastrophic damage to the film and upheaval of the surface were observed by SEM and AFM.

Ga$_2$O$_3$ has attracted significant interest in power electronics [1], solar-blind detectors [2], and high temperature electronics [3] due to its ultrawide bandgap and high breakdown electrical field. Ga$_2$O$_3$ is polymorphic, with the orthorhombic-structured $Ga_2O_3$ ($β$-$Ga_2O_3$) being the thermodynamically stable form [4]. With the largest bandgap of ~5.3 eV [5, 6] among all the polymorphs, corundum-structured $Ga_2O_3$ ($α$-$Ga_2O_3$) is a promising alternative material for applications in high-power electronics, as indicated by its higher Baliga figure of merit (BFOM) than $β$-phase $Ga_2O_3$[7]. $α$-$Ga_2O_3$ could be grown on isomorphic sapphire ($α$-$Al_2O_3$), of which large-scale wafers are commercialized at a reasonable price. Currently, conductivity control of n-type doping (mainly achieved by Sn or Si dope) $α$-$Ga_2O_3$ could be grown by mist-chemical vapor deposition (mist-CVD)[8, 9], pulsed laser deposition (PLD) [10], halide vapor phase epitaxy (HVPE) [11], and suboxide molecular-beam epitaxy (S-MBE) [12].

Moreover, $α$-$Ga_2O_3$ is able to form alloys with other corundum-structured oxides, such as III-oxides $α$-$Al_2O_3$ and $α$-$In_2O_3$, offering a bandgap energy tunable in wide range of 3.7-8.6 eV[13, 14]. Specifically, $α$-$(Al_xGa_{1-x})_2O_3$ can be grown without a compositional limitation[15] compared to $β$-$(Al_xGa_{1-x})_2O_3$. $(Al_xGa_{1-x})_2O_3$ covers 5.3-8.6 eV bandgap range and is predicted to exhibit one of the highest theoretical BFOM among ultra-wide-bandgap semiconductors. Currently, $α$-$(Al_xGa_{1-x})_2O_3$ has been introduced as a buffer layer in $α$-$Ga_2O_3$-on-sapphire based devices to reduce the lattice mismatch between $α$-$Ga_2O_3$ and sapphire substrate, which could improve $α$-$Ga_2O_3$ crystalline quality and hence device performance [16]. For example, benefited from the improved epitaxial quality with an $α$-$(Al_{0.24}Ga_{0.76})_2O_3$ intermediate layer, a smaller dark current and a higher responsivity are achieved in $α$-$Ga_2O_3$ based solar-blind detectors[17]. Similarly, due to the improved crystalline quality using the graded $α$-$(Al_xGa_{1-x})_2O_3$ layer, a larger breakdown voltages and forward current density are obtained in an $α$-$Ga_2O_3$ Schottky diodes[18]. Moreover, band-offset engineering in $α$-$(Al_xGa_{1-x})_2O_3$/$Ga_2O_3$ and progress in donor doping and heterostructure growth provide the materials basis for high electron mobility transistors (HEMT) to further improve $Ga_2O_3$ devices' performance[19-23]. Although modulation-doped $α$-phase HEMT remains less established in the open literature, modulation-doped $β$-$(Al_xGa_{1-x})_2O_3$/$Ga_2O_3$ heterostructures for HEMT have been demonstrated[24, 25]. Therefore, more effort is required to investigate the properties of $α$-$(Al_xGa_{1-x})_2O_3$/$Ga_2O_3$ heterostructures.

As metastable phase, one challenge of $α$-$(Al_xGa_{1-x})_2O_3$/$Ga_2O_3$ heterostructures is its thermal stability. Moreover, first principles calculations find that the thermal conductivity of $α$-$Ga_2O_3$ is low (8.0-11.6 W/mK) [26]. High temperatures environments occur during device fabrication processes and applications. Ion implantation is attractive for $Ga_2O_3$ device since it provides conductivity control in the lateral directions [27]. Implanted dopants require high temperature annealing to activate the carrier. Specifically, anneal temperature up to 900-1000 °C is required in $β$-$Ga_2O_3$ with Si-ion implantation doping [28]. Other processes such as ohmic contact also involve high temperature anneals [29, 30]. Besides, under a high-power and high-frequency operation, $Ga_2O_3$ based electronics faced an extremely high-power density and high-power loss. They lead to a high channel temperature that could be higher than 180 °C characterized by micro-Raman in a $α$-$Ga_2O_3$ MOSFET [31] and higher than 330 °C characterized by infrared thermal imaging in a $β$-$Ga_2O_3$ MOSFET [32].

The thermal stability of $α$-$Ga_2O_3$ and $α$-$(Al_xGa_{1-x})_2O_3$ have been studied both in theory [33] and experiment [34-39]. The growth method includes mist-CVD [34, 35, 37], ALD [36], MBE [38], and HVPE [39]. Previous study observed that the phase transformation temperature of $α$-$Ga_2O_3$ is dependent to thickness of $α$-$Ga_2O_3$ [37] and that of $α$-$(Al_xGa_{1-x})_2O_3$ is dependent to Al concentration [35]. However, no studies have investigated the thermal stability and phase transformation of the α-(Al$_x$Ga$_{1-x}$)$_2$O$_3$/Ga$_2$O$_3$ heterostructures. Moreover, most works do not focus on conductive α-Ga$_2$O$_3$ or α-(Al$_x$Ga$_{1-x}$)$_2$O$_3$ thin films, which are essential for application in power electronics.

This Letter reports the investigations of the thermal stability and phase transformation of conductive α-(Al$_x$Ga$_{1-x}$)$_2$O$_3$/Ga$_2$O$_3$ heterostructure on c-plane sapphire ((0001)) substrates using in situ high temperature X-ray diffraction (HT-XRD) up to 1000°C, including 2θ-ω scan, rocking-curve (RC), and reciprocal space mapping (RSM). As shown in Figure 1(a), 1-μm-thick α-Ga$_2$O$_3$ film followed by 100 nm α-(Al$_x$Ga$_{1-x}$)$_2$O$_3$ was grown on sapphire by mist-CVD [40]. For α-Ga$_2$O$_3$, gallium acetylacetonate (Ga(acac)$_3$) was dissolved in deionized water (DI water) with a small amount of HCl to prepare the solution. The solution was atomized using a 1.7 MHz transducer, and the mist was carried into a CVD reactor at 450 °C by N$_2$ carrier gas. Under these conditions, the growth rate was approximately 17 nm/min, and an α-Ga$_2$O$_3$ layer with a thickness of ~1 μm was obtained after 60 min. For α-(Al$_x$Ga$_{1-x}$)$_2$O$_3$, aluminum acetylacetonate (Al(acac)$_3$) and NH$_4$F were added in deionized water (DI water) with a small amount of HCl to prepare the solution for Al source and F dopant, respectively. Except from the n-type doping, F dopant element also provide better thermal stability in both electrical and structural properties in α-Ga$_2$O$_3$ than undoped or Sn-doped α-Ga$_2$O$_3$ as reported in our previous publication [40]. The solution was atomized using the same transducer, and the mist was carried into a CVD reactor at 470 °C by O$_2$ carrier gas. According to the XRD shown in Figure 1(b) and Bragg's law

$$2d\sin\theta = n\lambda \qquad (1)$$

where $\lambda = 1.5405$ Å for Cu K$\alpha_1$ radiation, $n = 6$ for (0006) XRD peak. The lattices constant of c-axis of α-Ga$_2$O$_3$, α-(Al$_x$Ga$_{1-x}$)$_2$O$_3$, and α-Al$_2$O$_3$ (sapphire substrate) can be extracted as 13.32 Å, 13.25 Å, and 12.90 Å, respectively. Furthermore, according to Vegard's law of $d_{(Al_xGa_{1-x})_2O_3} = xd_{Al_2O_3} + (1-x)d_{Ga_2O_3}$, which has been verified by experiment [41, 42], the Al concentration could be expressed as:

$$x = \frac{d_{Ga_2O_3} - d_{(Al_xGa_{1-x})_2O_3}}{d_{Al_2O_3} - d_{(Al_xGa_{1-x})_2O_3}} \qquad (2)$$

Even though the lattice constant can be affected by the strain from the substrate, through inserting the extracted lattices constant, the Al composition could be approximately calculated at around $x \approx 0.16$. Electrical properties (including mobility and carrier concentration) of the films were also characterized by Hall measurements (Lakeshore Fast Hall) at room temperature by using the van der Pauw method. Indium was applied for Hall measurements and then removed before XRD measurement. As shown in Figure 1(c), a relatively high mobility and carrier concentration has been achieved compared to other reports, which are 28 cm$^2$/(V·s) and 1.4×10$^{20}$ cm$^{-3}$, respectively.

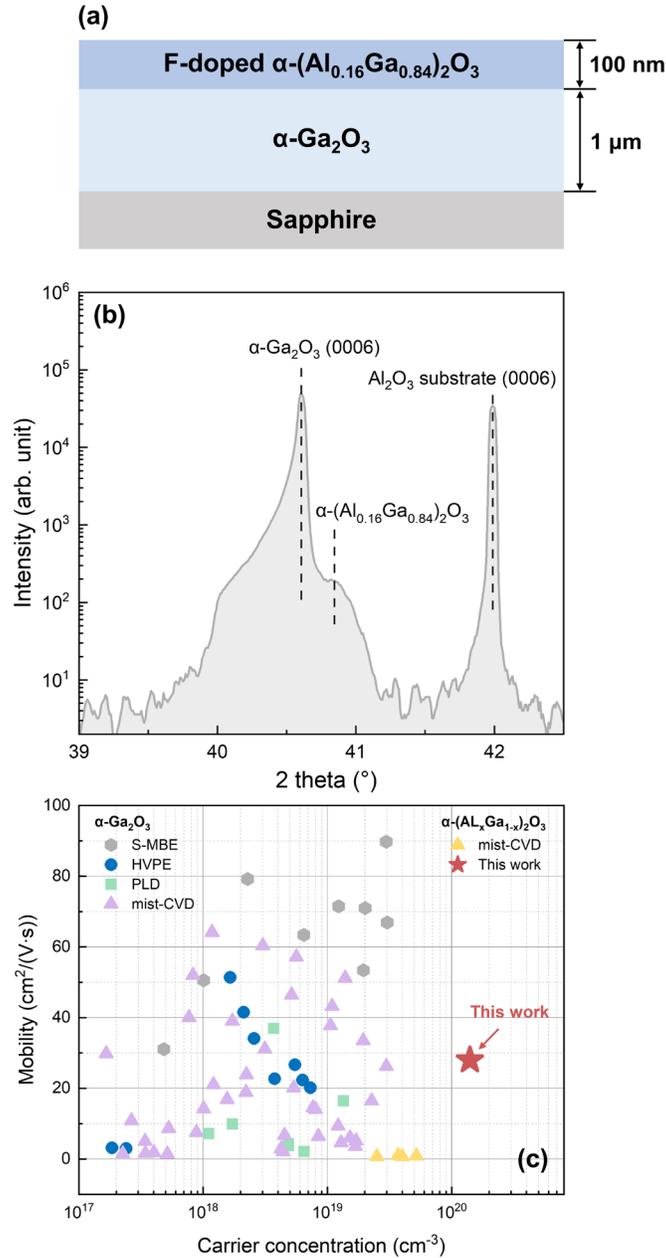

**FIG. 1** (a) Schematic of α-(Al$_x$Ga$_{1-x}$)$_2$O$_3$/Ga$_2$O$_3$ heterostructures on sapphire substrate. (b) XRD 2θ-ω scans at room temperature. (c) Comparison of mobilities of the doped α-Ga$_2$O$_3$ or α-(Al$_x$Ga$_{1-x}$)$_2$O$_3$ grown by S-MBE [12], HVPE [11], PLD [10], and mist-CVD [8, 9, 43, 44], as a function of mobile carrier concentration.

In-situ high-temperature X-ray diffraction (HT-XRD) measurements were performed using a Bruker D8 Discover diffractometer equipped with an Eiger 2R 250k detector and an Anton Paar DHS 1100 heating stage. During heating, the sample was enclosed by a graphite dome to protect the instrument from thermal radiation and to improve temperature uniformity. To evaluate the thermal stability and possible phase transformation of the α-$(Al_{0.16}Ga_{0.84})_2O_3/Ga_2O_3$ heterostructure, HT-XRD $2\theta$-$\omega$ scans, rocking curve (RC) scans, and reciprocal space mapping (RSM) were carried out in air atmosphere from 25 °C to 1000 °C, followed by a final scan after cooling back to 25 °C. A finer temperature step was applied between 500 °C and 600 °C to better capture potential onset of phase transformation. The temperature ramp rate between setpoints was 300 °C min$^{-1}$ up to 500 °C and reduced to 10 °C min$^{-1}$ above 500 °C. At each set temperature, the sample was held for more than 40 min in total, including the stabilization delay and acquisition time for the $2\theta$-$\omega$, RC, and RSM measurements. After the 1000 °C measurement, the sample was cooled from 1000 °C to ~27 °C while maintaining the graphite dome in place, with a cooling duration of approximately 20 min.

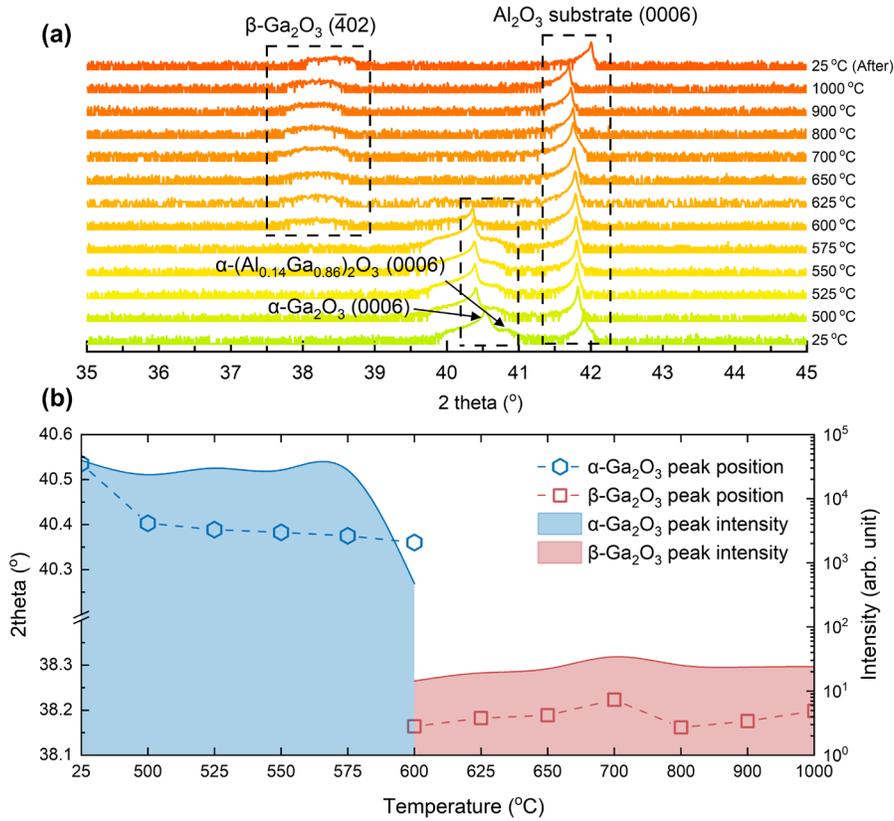

**FIG. 2.** (a) Temperature-dependent $2\theta$-$\omega$ XRD scans of the α-$(Al_{0.16}Ga_{0.84})_2O_3/Ga_2O_3$ sample measured from room temperature to 1000 °C. (b) Extracted peak positions and intensities of the α-$Ga_2O_3$ and β-$Ga_2O_3$ diffraction peaks as a function of temperature.

Figure 2(a) shows the temperature-dependent $2\theta$-$\omega$ XRD scans of the α-$(Al_{0.16}Ga_{0.84})_2O_3/Ga_2O_3$ heterostructure measured from 25 °C to 1000 °C. At room temperature (25 °C), two dominant diffraction peaks located at approximately 40.5° and 41.6° can be observed, corresponding to the α-$Ga_2O_3$ (0006) reflection and the sapphire substrate peak, respectively. As discussed above, a weaker peak appears on the high-angle side of the α-Ga$_2$O$_3$ (0006) peak, which is assigned to the α-(Al$_{0.16}$Ga$_{0.84}$)$_2$O$_3$ layer. With increasing temperature, both the α-Ga$_2$O$_3$ (0006) peak and the sapphire peak exhibit a continuous shift toward lower 2θ angles, indicating lattice expansion caused by thermal expansion. Notably, when the temperature reaches approximately 575 °C, a weak feature emerges near 38.2°, which is close to the expected position of the β-Ga$_2$O$_3$ ($\bar{2}$01) reflection. Upon further heating to 600 °C, the β-Ga$_2$O$_3$ ($\bar{2}$01) peak becomes significantly pronounced while the intensity of the α-Ga$_2$O$_3$ (0006) peak decreases, suggesting that the α to β phase transformation initiates. When the temperature increases to 625 °C, the α-Ga$_2$O$_3$ (0006) peak nearly disappears, and the diffraction pattern is dominated by the substrate peak and a broad β-Ga$_2$O$_3$ feature which indicates that the transformed β phase is mainly polycrystalline rather than epitaxially aligned. Further heating up to 1000 °C does not introduce additional phase changes, and the film remains in the β-Ga$_2$O$_3$ phase after cooling, consistent with the higher thermodynamic stability of β-Ga$_2$O$_3$.

Figure 2(b) summarizes the evolution of the peak position and peak intensity extracted from the temperature-dependent 2θ-ω scans, including the α-Ga$_2$O$_3$ (0006) and β-Ga$_2$O$_3$ ($\bar{2}$01) reflections. Before 600 °C, the α-Ga$_2$O$_3$ (0006) peak remains the dominant film diffraction feature, and its 2θ position continuously shifts toward lower angles with increasing temperature, consistent with lattice thermal expansion. In this temperature range, the α-Ga$_2$O$_3$ peak intensity does not show a strong monotonic decrease, indicating that the epitaxial α-phase remains largely preserved prior to the phase transformation. When the temperature reaches approximately 600 °C, a pronounced reduction in the α-Ga$_2$O$_3$ (0006) intensity is observed, accompanied by the emergence and rapid growth of the β-Ga$_2$O$_3$ ($\bar{2}$01) peak, suggesting the onset of the α to β phase transformation. After the transformation, the β-Ga$_2$O$_3$ ($\bar{2}$01) peak becomes the main Ga$_2$O$_3$-related reflection aside from the sapphire substrate peak. However, its intensity remains significantly lower, and the diffraction peak is broadened, implying that the transformed β-Ga$_2$O$_3$ is mainly polycrystalline rather than a high-quality epitaxial phase. In an ideal single-crystalline material, the coherent Bragg intensity is expected to decrease with increasing temperature due to enhanced atomic thermal vibration, which can be described by the Debye-Waller factor through the temperature-dependent scattering factor as[45]:

$$f = f_0 e^{-M} \qquad (3)$$

Here, $M$ increases with temperature and can be expressed as

$$M = 8\pi^2 \bar{u}^2 (\sin\theta/\lambda)^2 \qquad (4)$$

where $\bar{u}^2$ represents the mean-square atomic displacement. Therefore, the diffracted intensity is expected to decrease as temperature increases. In contrast, the β-Ga$_2$O$_3$ ($\bar{2}$01) peak shows an overall intensity increase after it first emerges during heating. This means that the β-Ga$_2$O$_3$ signal evolves dynamically during the HT-XRD process, consistent with progressive nucleation and grain growth of β-Ga$_2$O$_3$. As the temperature increased, the volume fraction of β-Ga$_2$O$_3$ increases continuously, resulting in a net gain in the diffracted intensity of the β reflection despite the Debye-Waller attenuation. Meanwhile, the β-Ga$_2$O$_3$ peak remains relatively broad compared with the original epitaxial α-Ga$_2$O$_3$ peak, suggesting that the transformed β-Ga$_2$O$_3$ is primarily polycrystalline with a distribution of grain orientations and/or residual strain. Furthermore, the apparent shift of the β-Ga$_2$O$_3$ peak position does not necessarily reflect pure thermal expansion behavior, because the measured feature likely includes contributions from multiple β-phase grains, overlaps with a broad background, and continuous redistribution of intensity within the broad peak envelope. Therefore, the intensity increases and peak evolution of the β-Ga$_2$O$_3$ reflection strongly supports an ongoing structural reconstruction and crystallization process during heating up to 1000 °C.

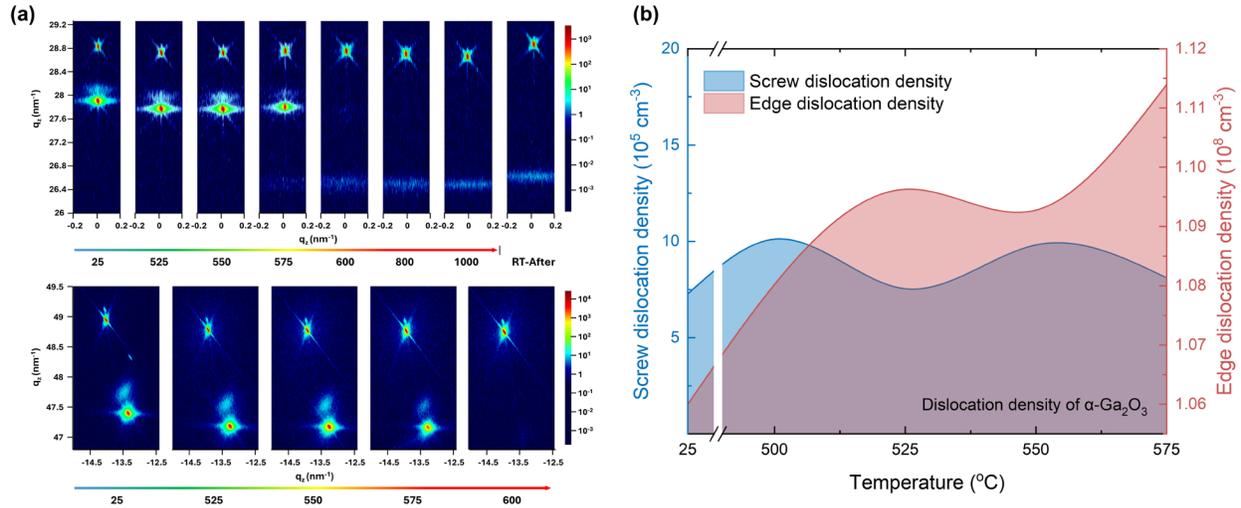

**FIG. 3.** (a) Temperature-dependent symmetric and asymmetric RSMs of the α-(Al$_{0.16}$Ga$_{0.84}$)$_2$O$_3$/Ga$_2$O$_3$ heterostructure measured at selected temperatures during heating. (b) Extracted screw and edge dislocation densities of the α-Ga$_2$O$_3$ layer as a function of temperature.

To further resolve the structural evolution during heating and to provide a clearer identification of the phase transformation observed in the 2θ-ω scans, symmetric and asymmetric RSMs were performed at selected temperatures, as shown in Figure 3(a). At 25 °C, the symmetric RSM exhibits three diffraction features distributed from high to low $q_z$, corresponding to the Al$_2$O$_3$ substrate, the α-(Al$_{0.16}$Ga$_{0.84}$)$_2$O$_3$ layer, and the α-Ga$_2$O$_3$ layer, respectively. The α-(Al$_{0.16}$Ga$_{0.84}$)$_2$O$_3$ peak is noticeably weaker than the substrate and α-Ga$_2$O$_3$ peaks, which can be attributed to its relatively small thickness and/or lower crystalline quality. With increasing temperature, no obvious degradation of the α-(Al$_{0.16}$Ga$_{0.84}$)$_2$O$_3$/Ga$_2$O$_3$ diffraction peaks is observed prior to ~575 °C, implying that the α-phase maintains stable crystallinity during heating below the phase-transition temperature. When the temperature reaches 575 °C, a faint diffraction feature emerges at the bottom of the symmetric map along with the intensity degrading of α-(Al$_{0.16}$Ga$_{0.84}$)$_2$O$_3$/Ga$_2$O$_3$ peaks, which can be assigned to the phase transition from α-phase to β-Ga$_2$O$_3$ ($\bar{2}$01). Upon further heating to 600 °C, the β-Ga$_2$O$_3$ ($\bar{2}$01) peak becomes substantially more pronounced, while the α-Ga$_2$O$_3$ peak becomes barely observable, suggesting that the α to β phase transformation becomes dominant in this temperature range. A consistent transformation behavior is also observed in the asymmetric RSMs, where the Al$_2$O$_3$, α-(Al$_{0.16}$Ga$_{0.84}$)$_2$O$_3$, and α-Ga$_2$O$_3$ diffraction peaks at (0$\bar{1}$10) are similarly distributed from top to bottom at low temperature. As the temperature increases, both α-phase peaks progressively lose intensity and become less distinguishable, further confirming that the *α*-phase structure becomes unstable during heating. Overall, the temperature-dependent RSM results provide direct reciprocal-space evidence that the *α*-phase heterostructure begins to transform at approximately 575 °C, and the diffraction response becomes dominated by the newly formed *β*-$Ga_2O_3$ phase at 600 °C.

Because an RSM records diffraction intensity as a function of both $\omega$ and $2\theta$, RCs can be extracted at each temperature, enabling evaluation of the corresponding full width at half maximum (FWHM). However, at elevated temperatures, the RC broadening reflects a combination of effects, including thermal vibrations, strain fluctuations, and potentially temperature-induced defect evolution. Despite these contributions, the *α*-$Ga_2O_3$ epilayer exhibits a more pronounced FWHM broadening than the substrate, indicating an enhanced contribution from film-specific structural disorder during heating. Therefore, the dislocation density was estimated from the measured FWHM values to quantitatively assess the evolution of crystalline disorder. Figure 3(b) shows the temperature dependence of the screw and edge dislocation densities in the *α*-$Ga_2O_3$ epilayer, calculated using the FWHM extracted from the symmetric RC and the asymmetric scan, respectively. The dislocation densities were estimated using the widely adopted relationship[46]:

$$\rho = \frac{\beta^2}{4.35\, b^2} \tag{5}$$

where $\rho$ is the dislocation density, $\beta$ is the FWHM, and $b$ is the magnitude of the relevant Burgers vector, which is approximate *c* lattice constant and *a* lattice constant of *α*-$Ga_2O_3$ for screw and edge dislocation density, respectively. As shown in Figure 3(b), the edge dislocation density remains consistently higher than the screw dislocation density throughout the entire heating process, indicating that the crystalline imperfection in *α*-$Ga_2O_3$ is dominated by in-plane mosaicist. This behavior is expected for heteroepitaxial *α*-$Ga_2O_3$ grown on sapphire, where lattice mismatch and thermal stress can preferentially introduce edge-type threading dislocations. Notably, the edge dislocation density shows a continuous increasing trend with increasing temperature, instead of remaining constant. This gradual increase suggests that the *α*-$Ga_2O_3$ layer undergoes progressive strain accumulation and defect evolution during high-temperature heating. In addition to intrinsic thermal vibration effects, the high-temperature environment may promote dislocation motion, defect multiplication, and strain relaxation, which can increase the mosaic twist and broaden the asymmetric diffraction response, leading to a higher extracted edge dislocation density. Furthermore, near the onset of the *α* to *β* phase transformation (around 575-600 °C), the structural instability associated with phase reconstruction can introduce additional lattice distortion and local strain fields, which further accelerates the degradation of crystalline coherence and contributes to the increased edge dislocation density. In contrast, the screw dislocation density shows a much weaker variation during heating, implying that the out-of-plane tilt component is less sensitive to thermal loading compared with the in-plane distortion.

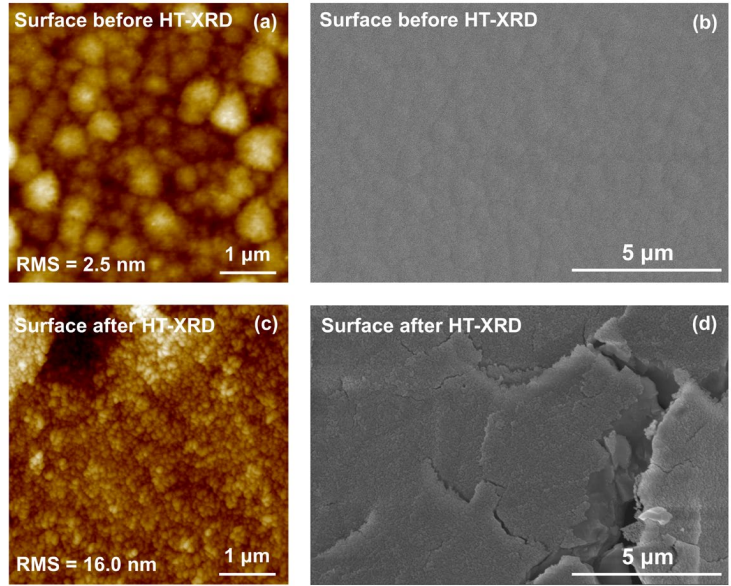

**FIG. 4.** Surface morphology of the sample before and after high-temperature XRD (HT-XRD). (a) AFM image of the as-grown surface with RMS roughness of 2.5 nm. (b) SEM image of the as-grown sample. (c) AFM image after HT-XRD heating. (d) SEM image after HT-XRD heating.

Figure 4 compares the surface morphology of the sample before and after the HT-XRD heating process using atomic force microscope (AFM) and scanning electron microscope (SEM). As shown in the AFM images in Figures 4(a) and 4(c), the as-grown surface exhibits a relatively uniform granular morphology with a low root-mean-square (RMS) roughness of ~2.5 nm, indicating a smooth and continuous film surface prior to heating. After the HT-XRD measurement, the surface becomes significantly roughened, with the RMS roughness increasing to ~16.0 nm. Consistent with the AFM results, the SEM images in Figures 4(b) and 4(d) show that the surface changes from a relatively smooth morphology before heating to a much rougher and discontinuous surface after heating. The post-heating SEM image also reveals the formation of large cracks and surface segmentation, indicating mechanical failure induced by thermal processing. This cracking behavior is likely associated with the combined effects of thermal stress, phase transformation, and microstructural reconstruction during the $\alpha$ to $\beta$ transformation. In particular, the transformation from the epitaxial $\alpha$-phase to the $\beta$-$Ga_2O_3$ phase is accompanied by a loss of coherent epitaxial registry and the formation of polycrystalline domains, which introduces localized strain and stress concentration, thereby promoting crack initiation and propagation. Therefore, the strong roughening and cracking observed after heating provide direct morphological evidence supporting the diffraction results that the film undergoes significant structural degradation and becomes dominated by a non-epitaxial $\beta$-$Ga_2O_3$ phase after high-temperature annealing.

In summary, thermal stability and phase transformation of conductive $\alpha$-$(Al_{0.16}Ga_{0.84})_2O_3$/$Ga_2O_3$ heterostructure on sapphire substrates were investigated via in situ HT-XRD and microscopy investigation. The heterostructure exhibited thermal stability to ~550-575°C before transforming to $\beta$-phase. The resulting $\beta$-$Ga_2O_3$ phase exhibited a broadened diffraction response, indicating that the transformed film is predominantly polycrystalline rather than a high-quality epitaxial layer. Reciprocal space mapping (RSM) further suggests that the crystalline imperfection in the $\alpha$-$Ga_2O_3$ layer is dominated by in-plane mosaicity. After the phase transformation of $\alpha$ phase to $\beta$ phase, catastrophic damage to the film and upheaval of the surface were observed by SEM and AFM. These results are an important reference for thermal stability of $\alpha$-$(Al_xGa_{1-x})_2O_3$/$Ga_2O_3$-based device.

## Acknowledgement

Botong Li, Bobby G. Duersch, Hunter Ellis, Imteaz Rahaman, and Kai Fu acknowledge the support from the University of Utah start-up fund and Research Incentive Seed Grant by the Price College of Engineering and the Vice President for Research (VPR) Office. This work made use of the Nanofab EMSAL shared facilities of the Micron Technology Foundation Inc. Microscopy Suite, sponsored by the John and Marcia Price College of Engineering, Health Sciences Center, Office of the Vice President for Research. In addition, it utilized the University of Utah Nanofab shared facilities, which are supported in part by the MRSEC Program of the NSF under Award No. DMR-112125. Acquisition of the Bruker D8 Discover system was made possible by the Air Force Office of Scientific Research under project number FA9550-21-1-0293. Shisong Luo, Lucas Lau, Cheng Chang, Zonghao Zhang, and Yuji Zhao are supported, in part, by the Nano & Material Technology Development Program through the National Research Foundation of Korea (NRF) funded by the Ministry of Science and ICT (RS-2024-00451173) and, in part, by CHIMES, one of the seven centers in *JUMP 2.0*, a Semiconductor Research Corporation Program by DARPA. The authors would like to acknowledge the use of XRD and AFM of Shared Equipment Authority at Rice University.